\documentclass{article}

\usepackage{arxiv}

\usepackage[utf8]{inputenc} 
\usepackage[T1]{fontenc}    
\usepackage{hyperref}       
\usepackage{url}            
\usepackage{booktabs}       
\usepackage{amsfonts}       
\usepackage{microtype}      
\usepackage{lipsum}
\usepackage{graphicx}
\graphicspath{ {./Figures/} }

\title{Beyond 5G Domainless Network Operation enabled by Multiband: Toward Optical Continuum  Architectures}

\author{
  \'{O}scar Gonz\'{a}lez de Dios\\
  Telefonica I+D, Spain\\
  \texttt{oscar.gonzalezdedios@telefonica.com} \\
  \And  
  Ramon Casellas\\
  CTTC/CERCA, Spain\\
  \texttt{ramon.casellas@cttc.es} \\
  \And  
  Filippo Cugini\\
  CNIT, Italy\\
  \texttt{filippo.cugini@cnit.it} \\  
  \And 
  Jos\'{e} Alberto Hern\'{a}ndez\\
  Universidad Carlos III de Madrid, Spain\\
  \texttt{josealberto.hernandez@uc3m.es} \\
}

\begin{document}

\maketitle

\begin{abstract}

Both public and private innovation projects are targeting the design, prototyping and demonstration of a novel end-to-end integrated packet-optical transport architecture based on Multi-Band (MB) optical transmission and switching networks. Essentially, MB is expected to be the next technological evolution to deal with the traffic demand and service requirements of 5G mobile networks, and beyond, in the most cost-effective manner. Thanks to MB transmission, classical telco architectures segmented into hierarchical levels and domains can move forward toward an optical network continuum, where edge access nodes are all-optically interconnected with top-hierarchical nodes, interfacing Content Delivery Networks (CDN) and Internet Exchange Points (IXP). This article overviews the technological challenges and innovation requirements to enable such an architectural shift of telco networks both from a data and control \& management planes.


\end{abstract}


\section{Introduction}


While mobile traffic currently accounts for approximately 9\% of the total IP traffic generated by the Internet, it is estimated that the upcoming massive deployment of 5G/IoT devices and services will double this value by the end of 2023, reaching 20\% of total IP traffic~\cite{cisco_vni}. Such a tremendous increase of connected devices and traffic is expected to push the underlying optical network infrastructure beyond its current limits. In fact, during the COVID-19 pandemic, many telecommunication operators (aka telcos) have reported extra traffic increase rates of 50\%~\cite{feldman_10.1145/3465212}. 

Besides, today's 5G transport is IP-centric supported by several optical networks spanning different domains (backbone, regional, metro, aggregation, access, etc). The photonic layer, often built with (Reconfigurable) Optical Add/Drop Multiplexers (ROADM) and Optical Line Terminals (OLT), is often used to interconnect IP routers among different domains in a hierarchical architecture of optically isolated networks, namely access, aggregation, metro and core. Indeed, electronic intermediate termination is currently present at the boundaries between network domains. 

Also, the traffic patterns are changing, and the network needs to migrate toward a data-center cloud ready infrastructure since the large majority of the traffic is no longer directed towards the IXP and Internet, but instead toward the caches in CDN scenarios. 

Innovation in optical networking solutions are required to provide sufficient and sustainable capacity capable of dealing not only with the expected increase in traffic and users/things, but also with the traffic pattern shifts expected in the next decade. 

One promising approach to increase bandwidth capacity without changing the optical infrastructure relies on Multi-Band (MB) transmission. Indeed, a number of research efforts, both from public and private funding, have realized the need to invest on MB networks as a way to squeeze the capacity of the already existing optical infrastructure, before deploying new fiber, as a cost-effective solution. 

MB expands the available capacity of optical fibres, by enabling transmission within S, E, and O bands, in addition to commercially available C and/or C+L bands. The total bandwidth available in MB is around 53 THz, which translates into more than 10x transmission capacity increase with respect to the C-band only. 

To realize MB networks, technology advances are required, both in the data and the control and management planes. The devices needed for MB data plane include new optical amplifiers, filter-less sub-systems, add/drop multiplexers, ROADMs, and transmission systems. These technology advances complement novel packet-optical white boxes using flexible MB sliceable bandwidth/bitrate variable transceivers (BVTs) and novel pluggable optics (e.g. of type Point-to-Multi-Point - PtMP such as XR, or 400 ZR/ZR+). The availability of MB transmission will also lead to a complete redesign of the end-to-end architecture, removing boundaries between network domains, reducing electronic intermediate terminations, and providing a network continuum between X-haul/access, aggregation, metro, and core segments. 

The control plane needs to be significantly extended to support MB elements and a ‘domain-less’ network architecture. 
The new control plane will extensively rely on physical layer abstraction, new MB impairment modelling, and pervasive telemetry data collection to feed AI/ML algorithms that will thus lead to a Zero-Touch Networking (ZTN) paradigm for MB networks. Ultimately this will result into a new, open network operating system deployed at the packet-optical white box.

This article attempts to shed light into the question on whether or not a new MB network infrastructure may allow telco architectures move toward an optical network continuum, where domain boundaries are removed and access nodes (central offices and mobile base stations) are all-optically interconnected to core and cache nodes bypassing classical electronic boundaries and aggregation stages. In addition, it shows what challenges and requirements are foreseen to fully realize such a radical architecture shift. 

The remainder of this work is then organised as follows: Section 2 is devoted to explain the telco architecture shift toward a domainless architecture. Section 3 overviews the main challenges in the data plane while section 4 does the same for the control and management planes. Finally, sections 5 and 6 show some preliminary assessment justifying the potential benefits of such a transition and the need to further investigate the challenges highlighted in this paper. Section 7 concludes this work with its main findings and conclusions.













\section{Telco hierarchical architecture}
\label{sec:arch}

\begin{figure*}[!t]
\centering
\includegraphics[width=\textwidth]{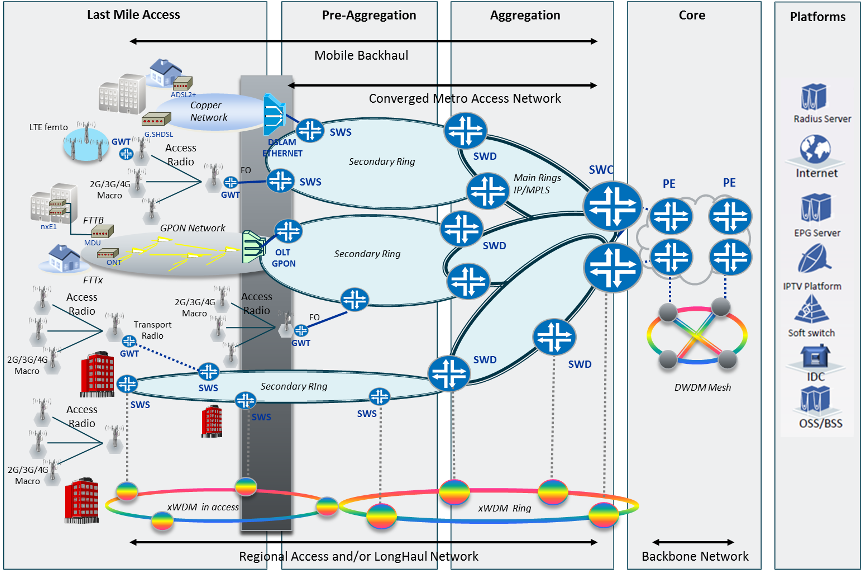}
\caption{Hierarchical and domain-segment architecture of telco operators}
\label{fig:telco_arch1}
\end{figure*}

Today's network architecture is built as a combination of IP/MPLS and optical WDM transport technologies, featuring a hierarchical structure spanning multiple domains, namely access, aggregation, metro and core domains (see Fig.~\ref{fig:telco_arch1}). These network domains span country-wide connectivity, including several metro/regional networks interconnected at city level and rural networks to reach small villages, both for fixed and mobile traffic.

\begin{figure}[!t]
\centering
\includegraphics[width=0.5\textwidth]{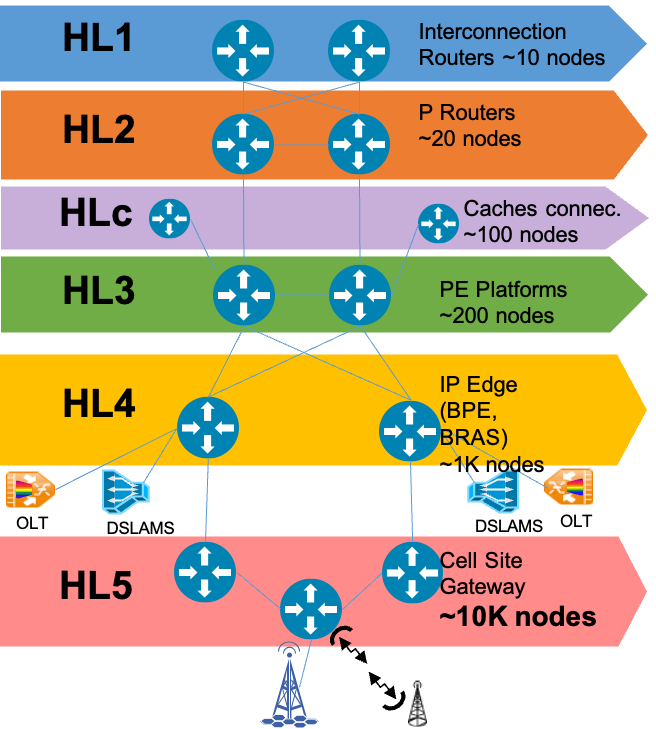}
\caption{Role and size of hierarchical levels}
\label{fig:telco_arch_hierarch}
\end{figure}

In a classical telco architecture, several Hierarchical Levels (HL) are often encountered with the following characteristics~\cite{Moreolo_jocn:21}:
\begin{itemize}
    \item HL5 typically comprise radio access points and mobile base stations, which are connected to cell site gateways, where traffic is encapsulated in L3VPNs to reach the packet core.
    \item HL4 nodes (SWS nodes in Fig.~\ref{fig:telco_arch1}) mostly comprise residential access, typically DSLAMs aggregating xDSL traffic and OLTs aggregating FTTx households.
    \item HL3 (SWD nodes in Fig.~\ref{fig:telco_arch1}) comprise a first metro node which collects the traffic from HL4 and HL5 nodes and performs IP grooming toward the next level.
    \item HL2 nodes (SWC nodes in Fig.~\ref{fig:telco_arch1}) often host MAN services for content distribution, such as CDN caching and IPTV services.
    \item HL1 nodes (PE nodes in Fig.~\ref{fig:telco_arch1}) comprise interconnection to the WAN.
\end{itemize}
The role and network size of each node is also summarised in Fig.~\ref{fig:telco_arch_hierarch}, where a simplified hierarchical architecture is shown.

Such high number of domains makes network operations complex and requires multiple O/E and E/O conversions and the traversal of intermediate IP routers. In addition to this, telco networks are evolving toward a data-center centric network, where most of the end-user traffic is terminated at the CDNs and OTT caches (Google, Facebook, Netflix, etc)~\cite{Moreolo_jocn:21}. The rise of datacenters spread across the telcos are growing exponentially, and the edge computing paradigm advocates for introducing many more small micro-data centers closer to the users, on attempts to provide ultra-low latency and improved user experience~\cite{hernandez_netsoft}. 

In spite of the introduction of ever-faster high-capacity channels (400~Gb/s and upcoming 800~Gb/s), the network infrastructure remains static at the switching level, where packets need to traverse multiple domains (IP and optical) and still perform several IP routing hops (typically between 4 and 6 hops, according to Telefonica's estimates). 

In this sense, multi-band networks provide a good opportunity to move from this rigid domain-based architecture into a flexible domainless architecture as shown in Fig.~\ref{fig:telco_opt_continuum} where HL4 nodes directly interconnect with HL2 or HL1 nodes bypassing intermediate HL3 nodes. In order to create a sustainable network continuum from access to core and DC, current rigid boundaries need to be broken. This optical continuum is expected to drastically reduce O/E/O needs and will allow to provide differentiated transparent connectivity services across network segments. 

To provide such comprehensive and ambitious approach, a number of research innovations in both the data and control \& management planes are required, namely:
\begin{itemize}
    \item A MB optical transport network infrastructure, based on specifically designed MB node architectures, MB switching, MB amplification, and MB transmission solutions, cost-effectively addressing the 10x bandwidth increase, providing open and fast programmability, and enabling an end-to-end continuum that drastically reduces electronic regeneration across network segments.
    \item A MB low-latency optical X-haul infrastructure exploiting pluggable-based PtMP heterogeneous solutions to enable massive Beyond-5G (B5G) and small cell deployments and their integration within the transport infrastructure and the support of QoS-guaranteed services (e.g., in case of mobility).
    \item A modular zero-touch control plane, able to orchestrate services across the end-to-end underlying infrastructure, jointly optimizing the packet and MB optical layers, while ensuring scalability and reliability, targeting zero-touch operations and relying on cluster-based distributed systems that synchronize state and exchange Machine Learning (ML) models.
\end{itemize}
The next sections further elaborate on such required innovations in both the data and control \& management planes toward an all-optical MB-based network continuum.


\section{Multi-band data plane: challenges from an Operator's perspective}

\begin{figure*}[!t]
\centering
\includegraphics[width=\textwidth]{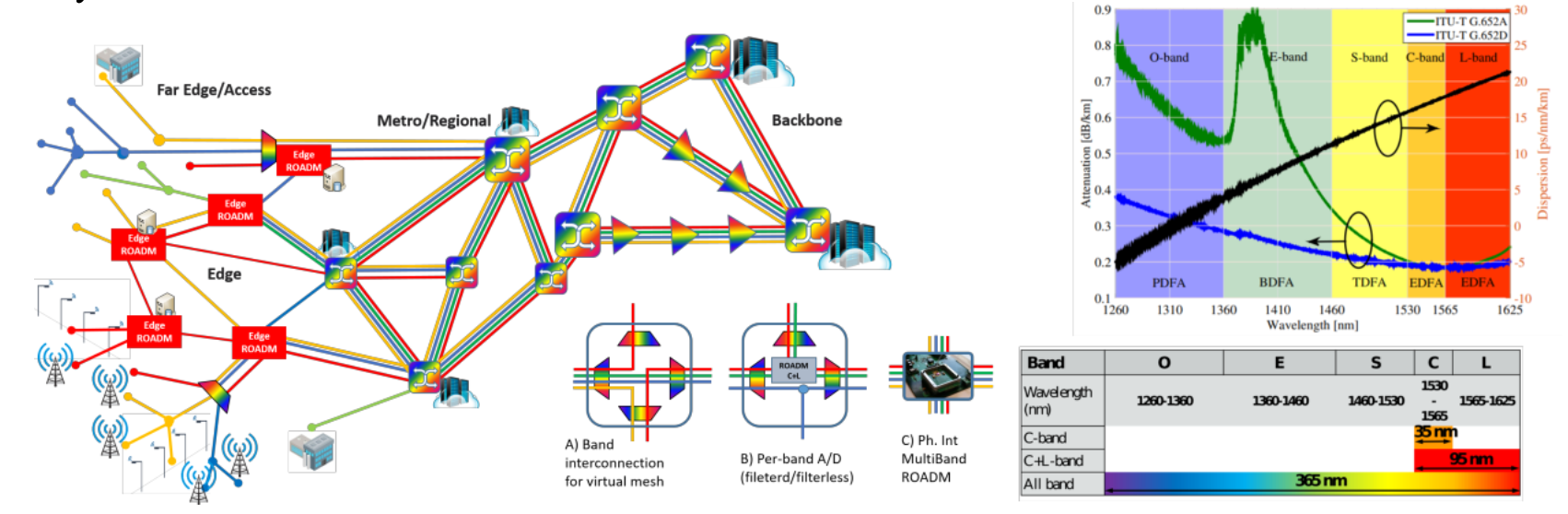}
\caption{MB network architecture with per-band differentiated reach and services (left); MB switching/ROADMs (bottom); MB considered spectrum with measured loss of G.652D fibre (right)}
\label{fig:telco_opt_continuum}
\end{figure*}

Fig.~\ref{fig:telco_opt_continuum} shows the proposed innovative data plane infrastructure based on MB technology to provide a network continuum from 5G access to data centers, with differentiated per-band optical transparency and services. Connectivity services over short distances will exploit the shorter wavelength range, where higher fibre attenuation is experienced and amplification systems may not be cost-effective~\cite{Ferrari:20}, adopting a pay-as-you-grow strategy. As shown in~\cite{Paolucci:20} a differentiated use of bands provides remarkable benefits to increase capacity while mitigating the complexity of additional hardware subsystems. For example, channels in the S band (yellow lines in Fig.~\ref{fig:telco_opt_continuum}) may be present on some links only, transparently crossing some intermediate nodes and domain boundaries, and being able to be added/dropped only in a few selected nodes. This way, seamless capacity scaling of the end-to-end network might be achieved, addressing the scaling disparities in the various optical network segments (access, aggregation, metro) while reducing the need for O/E/O regeneration at domain borders.

MB will leverage on 365 nm spectrum resources, from 1260 to 1625 nm, i.e. 53.4 THz, more than 12-fold the C-band spectral range. The orange area in Fig.~\ref{fig:telco_opt_continuum} (right) depicts the range of the currently used C-band. 

To fully take advantage of MB resources in Telco networks and efficiently implement the aforementioned data plane infrastructure with differentiated per-band optical transparency, a number of challenges need to be specifically addressed at the data plane level.

First, new cost-effective devices need to become commercially available. These include amplifiers, transmission modules, and optical switching elements. 

MB amplification has been demonstrated for transmission bandwidths up to around 100 nm. In most cases, hybrid amplification leveraging on EDFA, Raman, and Semiconductor Optical Amplifiers (SOA) were used. In~\cite{ham20}, the maximum experimented Raman amplification bandwidth was 83 nm. Such demonstration was done without exploiting mature control techniques (e.g., no gain control). Novel S-band optical amplifiers (Thulium-Doped Fiber Amplifier - TDFA) are nowadays commercially available at a cost significantly higher than EDFAs and with lower performance (e.g., non-flat noise figure of around 7 to 8 dB). 

Different transceiver solutions have been proposed in the literature towards supporting the fast evolution of current optical networks. In this context, point-to-multipoint coherent transceivers, also called sliceable bandwidth variable transceivers (S-BVTs), have assumed an especially relevant role in the research community during the last years, following the roadmap of optical networks. Current S-BVTs consist of a set of (white box) BVTs that transmit multiple slices/signals within the C band. The S-BVT enables disaggregation of the different optical components (BVTs) and also of hardware (HW) from software (SW)~\cite{nad20}. Several transceiver designs and configurations have been presented in the literature targeting different network segments. Many digital signal processing (DSP) and front-end transceiver options have been covered, also including approaches relaying on photonic technologies, according to the specific target application~\cite{Sva19}.
Transmitter/receiver modules on different bands have been so far designed mainly for low cost and short reach access solutions. That is, significant improvements are needed to make them cost-effective for medium/long term and dense multi-band WDM. 

In terms of MB node solution, only prototypes at the research level have been demonstrated including some switching capabilities or wavelength blockers operating across multiple bands (e.g.~\cite{Kra21}).  Significant work is needed to efficiently implement node solutions enabling differentiated per-band optical switching and bypass or blocking.

Besides data plane components and system solutions, advanced MB modeling needs to be efficiently implemented for the design of production-ready MB networks as well as to be encompassed within enhanced spectrum assignment algorithms, accounting for target reach and detailed in/cross-band impairments. 
Concerning transmission impairments like amplified spontaneous emission (ASE) and four-wave mixing (FWM), the use of MB networks notably changes the non-linearities, such that further extensions are required. The impact of these impairments in network capacity is assessed by means of Routing, Modulation and Spectrum Assignment (RMSA) algorithms ~\cite{val19}, albeit the studies are addressing C-band only. Moreover, ML techniques have been proposed to improve the accuracy of these estimations~\cite{sar19} as an independent approach.

\section{Zero-touch Control and Service Orchestration}

The control plane represents the intelligence of a network; in this sense, next-gen control planes aim at the design and development of zero-touch control and self-management of the underlying end-to-end data plane infrastructure and for the dynamic provisioning, management and orchestration of services, relying on dynamic planning covering the joint orchestration of IT and Network equipment. At present, network operators deploy optical nodes provided as an end-to-end closed solution, with proprietary device configuration interfaces. However, there is a trend to change this status-quo toward open optical networks, which require: (1) interoperability through standard application programming interfaces and (2) SDN Control Plane capabilities, based on standard models, to glue multivendor solutions~\cite{commag_disagg}.

\subsection{Innovation on control of MB transport infrastructure}

The control plane needs to be significantly extended to support MB elements and a ‘domain-less’ network architecture. New open and standard YANG models and interfaces for the forthcoming MB devices (i.e. ROADMs, amplifiers, pluggables and S-BVTs) need to be designed and validated. The optical MB infrastructure needs to be seamlessly integrated with the packet layer, thanks to the removal of standalone transponders and the extensive use of packet-optical white boxes, equipped with pluggable modules, P4-based packet forwarding programmability~\cite{pao21en}, and embedded monitoring and self-diagnosis/adaptation capabilities based on local control loops. To this extent, new open packet-opto operating systems are necessary~\cite{sca21}.

Specific procedures that enable self-management, self-diagnosis, and self-optimization through adaptation procedures of such devices are required. These include automatic inventory management, MB impairment-aware path computation, MB topology abstractions, and automatic identification of candidate pre-validated configurations. All these procedures need to be effectively supported by AI/ML and telemetry-based monitoring (with scenario dependent volume and rate of data) providing predictive capabilities for innovative SDN-based autonomous and ZTN.

\subsection{Innovation in Open Packet-Opto Operating System}

\begin{figure*}[!t]
\centering
\includegraphics[width=0.8\textwidth]{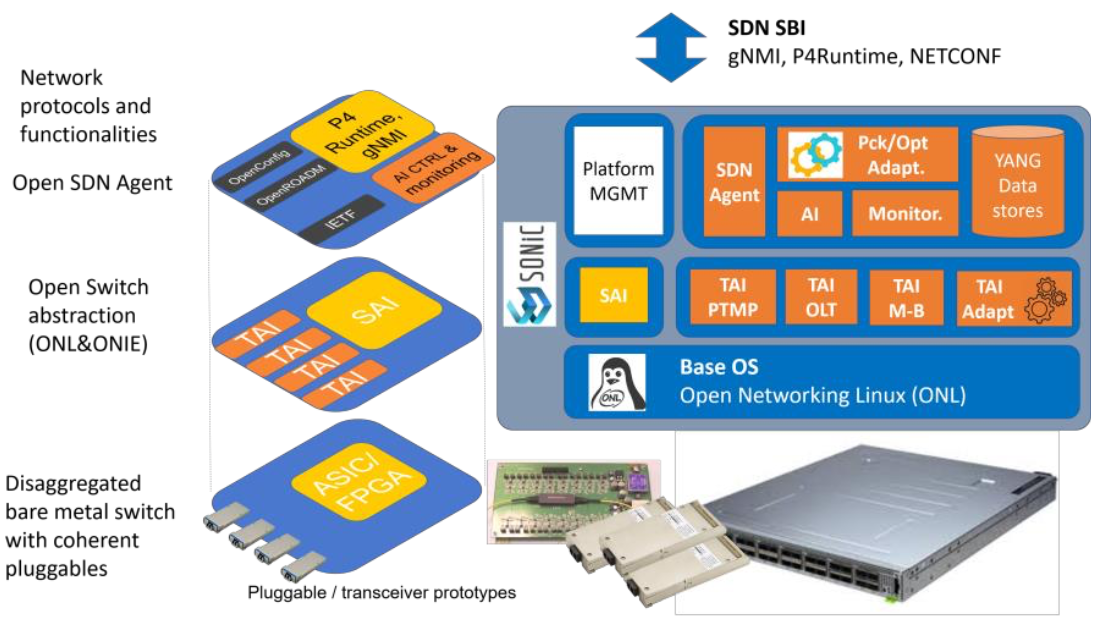}
\caption{Open packet-opto operating system (in orange the novel SW components).}
\label{fig:baremetal}
\end{figure*}

Fig.~\ref{fig:baremetal} shows the architecture of a possible example of white box open operating system, relying on emerging initiatives from open hardware. A disaggregated bare metal switch with coherent pluggable interfaces and FPGA-controlled novel MB transmission prototypes can be abstracted via a base OS (e.g. Open Networking Linux - ONL). 

Then, specifically designed Transceiver Abstraction Interfaces (TAI) have to be developed to support novel pluggable modules for MB, PtMP digital subcarrier multiplexed (DSCM), and OLT modules. Self-adaptation of transceiver parameters (e.g. auto-tunability, operational mode adaptation, etc.) will have to be efficiently supported, also relying on finite state machines for proactive enforcement. At the upper layer, extensive monitoring will be needed, including optical parameters collected via TAI, metadata from the P4 pipeline, and in-band telemetry. Furthermore, AI-based local elaborations with feature extraction at the data plane will feed a novel module providing local control loops and autonomous adaptations of the whole node functionalities (e.g. fast recovery/rescheduling of low latency flows toward different bands/pluggables). The traditional local SDN agent will have to be extended to provide both optical and packet continuous interaction with the new SDN domain-less control plane. The OS should rely on existing modules that are part e.g. of the SONiC OS, a reliable platform supporting, besides legacy protocols (e.g. Border Gateway Protocol - BGP), effective integration with the Kubernetes platform. The integrated solution will create a powerful open packet-optical operating system with large potential impact, specifically tailored to meet network operators’ needs.

\subsection{Innovations in SDN controller architecture}

SDN controllers tend to be monolithic and cover mainly a single optical domain. However, there is a clear need to design and develop new controller architectures (e.g. using a micro-services approach combining multiple functional elements); to consider new control plane deployment models (where such functional elements are deployed as virtualized functions or containers);  to tightly integrate configuration, control, and monitoring and to address scalability issues by deploying multiple controllers, whose arrangement is dynamic, involving high volume communications and new interfaces between controllers. Gigabit/s data links supporting control plane communications are becoming available and this allows improved distributed systems, fast state synchronization between controllers, huge volume monitoring, and real time telemetry and closed control loops based on AI/ML. This is realized via collaborative knowledge-management based on data and model exchange among node agents, as well as between the network controller and node agents encompassing centralized vs distributed vs coordinated autonomous decision making. Modern SDN architectures shall also address infrastructure sharing and network virtualization/slicing. Indeed, MB is an opportunity for efficient network sharing at the optical layer, including the virtualization of access networks (vPON, vOLT) towards the B5G Radio Access Network (RAN), along with new business models related to multi-tenancy.

\subsection{Innovations regarding end-to-end orchestration}

As stated, there is a need to address multi-domain SDN to support end-to-end orchestration of highly dynamic services, defining architectures that span dynamic network segments that today correspond to the access, metro and part of the core and allow seamless integration with the RAN, the integration with computing domains and the "federation" of operations and vendor islands, which are heterogeneous in nature. 

Dedicated Network Planning subsystems and framework will assist the e2e orchestration and the SDN controllers in the decisions regarding placement and networking resource allocation divided in two blocks: In-Operation Provisioning assists the major components in the Infrastructure Control and Service Orchestration platform in the form of decision-making for resource allocation both in terms of IT and networking infrastructure

\section{Potential impact and benefits for Telco networks}

This section aims to provide a preliminary comparison in terms of  potential benefits for telco architectures based on MB networks with respect to existing solutions. 

As a baseline for comparison, we assume a classical telco architecture using N parallel fiber systems exploiting the C-band only, allowing 40 channels at 400~Gb/s per channel in a hierarchical architecture, where traffic is originated at the HL4s and directed toward the HL1/2 nodes, using HL3 nodes as intermediate nodes where IP grooming is performed (see section~\ref{sec:arch}) with $\eta$ oversusbscription ratio (typically $\eta$ is 1/2 or 1/3 at the metro/core, and 1/10 in the access)~\cite{rafa_oversubs,Oversubs_comm_letters}.

In this benchmark scenario, let us consider a hierarchical architecture where the lower, intermediate and upper hierarchical layers comprise $H_4$, $H_3$ and $H_{1/2}$ nodes respectively. In this setting, traffic is sourced from layers HL4 and HL3 and directed toward the closest HL1/2 node; in addition, intermediate HL3 nodes collect and aggregate traffic from a number of source HL4 nodes and perform grooming at the IP layer and forward this traffic to the closest HL1/2 node, using some over-subscription parameter $\eta$ at the grooming stage. In this setting, the number of 400G transponders needed at the HL4, HL3 and HL1/2 nodes to satisfy the traffic sourced at the HL4 nodes are:
\begin{eqnarray}
HL4\quad  N_{tx,HL4} & = & \Big\lceil\frac{A_4}{400G} \Big\rceil H_4  \\
HL3\quad  N_{tx,HL3} &= & \Big\lceil\frac{A_4}{400G} \Big\rceil H_4 + \Big\lceil\frac{H_4}{H_3} \frac{\eta A_4}{400G}\Big\rceil H_3  \\
HL12\quad  N_{tx,HL12} &=& \Big\lceil \frac{H_4}{H_3} \frac{\eta A_4}{400G} \Big\rceil H_3 
\end{eqnarray}
where $A_4$ refers to the average source traffic originated per HL4 node. Also, the intermediate HL3 layer assumes $1/2$ over-subscription ratio. In total, the number of 400G transponders needed in this scenario can be approximated as:
\begin{equation}
    N_{tx,IPoWDMgrooming} \approx (1+2\eta) \cdot \frac{A_4}{400G} H_4
\end{equation}

In an optical-continuum scenario where intermediate HL3 nodes are avoided, the number of 400G transponders required to interconnect HL4s directly to HL1/2s are:
\begin{eqnarray}
HL4\quad  N_{tx,HL4} & = & \Big\lceil \frac{A_4}{400G} \Big\rceil H_4  \\
HL12\quad  N_{tx,HL12} &=& \Big\lceil \frac{A_4}{400G} \Big\rceil H_4  
\end{eqnarray}
for the same $A_4$ offered traffic, thus yielding a total of approximately:
\begin{equation}
    N_{tx,IPoWDMcont} \approx 2 \cdot \frac{A_4}{400G} H_4
\end{equation}

As an example, consider a large MAN with 200 HL4 nodes, 40 intermediate HL3 nodes and 5 HL1/2 nodes providing connectivity with CDN contents and IXP, and further assume a traffic demand below 400G per HL4 node, say 300G peak offered traffic. This amount corresponds to an HL4 node serving 10,000 customers/things (both fixed and mobile) offering approximately 30~Mb/s offered traffic per customer/thing, in line with the estimates of~\cite{Hernandez_estimates}.

In this scenario, each HL4 node has got one 400G transponder to its intermediate HL3; the intermediate HL3 nodes have got another transponder to each HL4. In addtion, there is one HL3 per 5 HL4 nodes, thus aggregating the traffic of $5\times 300G =1,500G$ but, thanks to the $\eta = 0.5$ over-subscription ratio, the HL3 needs to provision capacity to only $750G$ toward the upper HL1/2 node, i.e. two 400G transponders, making a total of seven 400G transponders per HL3 node (5 in the downstream direction toward 5 HL4 nodes plus another 2 toward its HL1/2 node). Finally, the HL1/2 nodes need a 1-to-1 mapping of transponders to the HL3, thus, the total number of transponders per hierarchical layer is:
\begin{itemize}
    \item At HL4, 1 transponder per node times 200 nodes = 200 transponders
    \item At HL3, 5x40=200 transponders interfacing the HL4s plus another 2x40 = 80 transponders interfacing the HL12, for a total of 280 transponders.
    \item At HL12, 5x16=80 transponders interfacing the HL3s
\end{itemize}
As shown, the total number of 400G transponders required in this scenario is 560 transponders. In the optical continuum scenario avoiding intermediate HL3 nodes, the number of transponders needed are 200 in the HL4 layer plus another 200 in the HL1/2 layer, accounting for a total of 400 transponders. This implies a total saving of  28.5\% transponders with respect to the classical layered architecture approach. 

In addition to providing a faster direct all-optical connectivity, another advantage of this architecture is the removal of expensive high-capacity IP routers at intermediate HL3 nodes. 
Following the cost-estimates of~\cite{cost_model_jocn} and updated in~\cite{hernandez_netsoft}, where the cost of a 400G transponder is 12 normalised Cost Units (CU) and a Large-size IP router costs 64 Cost Units, then the total savings including transponders and intermediate IP Routers could rise up to 35-40\% approximately, since the layered architecture accounts for $580\times 12+20\times 60 = 7728~CU$ while the optical continuum is only $400\times 12 = 4800~CU$.

Although more detailed and extensive techno-economic analysis needs to be conducted, the above example shows the potential cost-savings of employing IP offloading of intermediate HL3 nodes, by interconnecting directly HL4 nodes with top-level HL1/2. Obviously, this approach requires more wavelengths available per fiber, which is only possible in MB networks (900 wavelengths MB vs only 80 wavelengths in C-band only~\cite{Ferrari:20}).









\section{MB pluggable S-BVT}

\begin{figure}[!htbp]
\centering
\includegraphics[width=\textwidth]{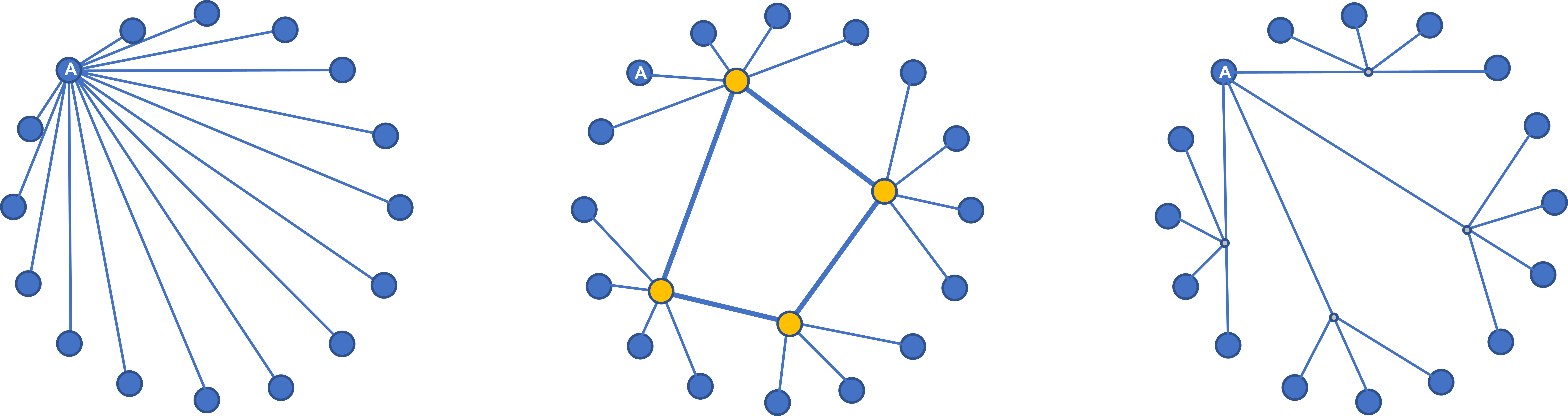}
\caption{Connectivity solutions based on: direct interconnection in full mesh (left); hierarchical electronic grooming (center); proposed direct interconnection through point-to-multipoint pluggable transceivers over MB in packet-optical nodes.}
\label{fig:mesh}
\end{figure}

This section provides a preliminary assessment of the potential benefits achievable by the introduction of PtMP coherent pluggable modules supporting MB transmission. Fig.~\ref{fig:mesh} shows three reference solutions to enable any-to-any connectivity at the packet level.

Fig.~\ref{fig:mesh} (left) shows a traditional approach based on direct connectivity among the considered $N$ network nodes. In the figure, node $A$ has a dedicated interface and a dedicated direct link (i.e. optical channel) toward any other node of the considered network. This approach suffers from two main scalability problems already for relatively limited values of $N$. First, it leads to a number \textit{N-1} of interfaces per router which is impractical to support. Second, the number of required channels for a full mesh optical interconnection rapidly exhausts the traditionally available spectrum in the C band.

For these two reasons, the traditional approach to enable any-to-any connectivity at the IP level leverages on a hierarchical infrastructure, shown in Fig.~\ref{fig:mesh} (center). In this exemplary case, node $A$ has a single interface and a single optical channel towards a transit router where traffic grooming is performed (yellow node in the figure). The transit router belongs to a hierarchical backbone which assures, through possible multi-hop interconnection, the required full mesh logical connectivity. This approach can be implemented guaranteeing limited use of interfaces per node and limited number of channels in the network, being able to cope with both node and C-band available resource. However, this approach is not free from limitations. The backbone routers are expensive and power-hungry electronic nodes. Furthermore, the end-to-end latency is increased due to the presence of intermediate hops and queuing stages, potentially subject to congestion issues.

Fig.~\ref{fig:mesh} (right) shows the proposed approach leveraging on novel PtMP coherent pluggable modules (i.e. pluggable S-BVT) operating in MB range and equipped within routers as packet-optical nodes. Such PtMP interfaces enable the implementation of direct any-to-any connectivity without requiring \textit{N-1} interfaces in the node. For example, a 400~Gb/s pluggable module could be implemented as a sliceable 4x100Gb/s solution towards 4 different nodes as in the figure. The second potential limitation in terms of number of required channels is solved by the availability of spectrum resources enabled by multiple bands. The presence of intermediate electronic grooming would be removed or strongly reduced leveraging on transparent optical switching/splitting solutions (gray nodes in the Fig.~\ref{fig:mesh} right). 

It is important to highlight that PtMP coherent pluggable modules are already commercially available (e.g., XR Optics) although operating in the traditional bands only and mainly targeting x-haul scenario. In this work, we envision the evolution to operate over MB and across metro-regional distances.

A possible implementation of MB PtMP pluggable modules in the previously considered benchmark scenario could consider to bypass HL3 nodes and interconnect HL4 directly to HL1/2 nodes. The number of PtMP pluggable following a 1:m fanout (i.e. 1:4 implies 4 x 100G) per layer follows:
\begin{eqnarray}
HL4\quad  N_{tx,HL4} & = & \Big\lceil \frac{A_4}{400G/m} \Big\rceil H_4  \\
HL12\quad  N_{tx,HL12} &=& \Big\lceil \frac{A_4}{400G/m} \Big\rceil H_4 \cdot \frac{1}{m}  
\end{eqnarray}
which, in total, the number of transceivers approximately follows:
\begin{equation}
    N_{tx,PtMP} \approx \frac{A_4}{400G/m} H_4 \cdot \left(1+\frac{1}{m}\right)
\end{equation}

Following the example in the previous section, where each HL4 offers 300G toward the closest HL1/2, we have that every HL1/2 concentrates the traffic of 200/5 = 40 HL4 nodes, i.e. 12,000G or 12T. In that case, the total number of needed PtMP pluggables is 200 modules for the HL4 nodes plus another 12T/400G = 30 modules per HL1/2, leading to a total of 200 + 5x30 = 350 PtMP modules. This number is significantly smaller than the 580 needed in the IPoWDM with IP grooming at the HL3 nodes, and also smaller than the 400 case in the optical continuum case with fixed 400G transponders. 

Also in this case, more detailed and extensive techno-economic analysis needs to be conducted to accurately assess the potential of novel PtMP pluggable modules over multi-bands.

\section{Summary}

This article overviews the concept of optical continuum architectures enabled by  multiband network technology. Indeed, emerging multiband techniques may allow up to 900 wavelengths per optical fibre using the S, O, E, C and L bands, thus providing a potential 10x increase in network capacity which can be exploited to providing direct lightpaths from 5G access/x-haul nodes toward the core and CDN nodes. This approach allows removing/reducing boundaries between domains (access, aggregation, metro) and further limiting the number of E/O and O/E terminations and IP routers at intermediate layers, with subsequent cost savings and simplified network operations. 

However, for this approach to become a reality, a number of technical challenges in both the data and control and management planes need to be addressed. Not only new MB devices need to be developed and matured, but also the necessary open interfaces and APIs need to evolve to allow SDN-based zero-touch open control. Doubtlessly, the next few years will witness a large number of research articles addressing some or many of these aspects toward multiband networks.



\section*{Acknowledgements and Funding} 

The authors would like to thank all B5G-OPEN Partners for the fruitful discussions and valuable contributions.

This work is partially funded through the European Commission B5G-OPEN project (grant no. 101016663), see \url{https://www.b5g-open.eu/}, last access February 2023, for further details.






\bibliographystyle{plain} 
\bibliography{multiband,PONs,costmodel,disaggregation,sample}

\end{document}